\documentstyle[12pt]{article}
\def\be{\begin{equation}}
\def\ee{\end{equation}}
\def\ben{\begin{displaymath}}
\def\een{\end{displaymath}}
\def\ba{\begin{array}{c}}
\def\bal{\begin{array}{l}}
\def\ea{\end{array}}
\def\p{\partial}
\begin{document}
\titlepage
.

\vspace{.35cm}

 \begin{center}{\Large \bf
 Mass-sign duality of cubic oscillators
  }\end{center}

\vspace{10mm}

 \begin{center}

 {\bf Miloslav Znojil}

 \vspace{3mm}
\'{U}stav jadern\'e fyziky AV \v{C}R,

250 68 \v{R}e\v{z}, Czech Republic

{e-mail: znojil@ujf.cas.cz}

\vspace{3mm}

\vspace{5mm}

%\today, imacu.tex

\end{center}

\vspace{5mm}

 \noindent
%\section*{Abstract}
In the light of the guiding methodical role of anharmonic
oscillators in field theory we communicate the observation that
their ``first nontrivial" ${\cal PT}-$symmetric cubic Hamiltonians
$H_\pm = p^2\pm m^2 {x}^2 +{\rm i}\,f\,{x}^3$ with opposite sign
of mass term are, up to a constant shift, isospectral.

\vspace{15mm}

\newpage

%\section{Introduction}

 \noindent
Quantitative description of quantum fields is mostly based on
perturbation theory where a free Lagrangean, say,
 \ben
 {\cal L}_{\rm free}= \frac{1}{2}\,\left [\p \phi(z)\right
 ]^2- \frac{1}{2}\,m^2\left [\phi(z)\right ]^2\,
 \een
is being complemented by a suitable interaction term ${\cal
L}_{\rm int}$ proportional to a ``small" coupling constant
$\lambda$. On a formal level, an enormous popularity of the
similar models with polynomial
 $
 {\cal L}_{\rm int} \sim \left [\phi(z)\right ]^3 + \ldots
 $
is based on their renormalizability as well as on the feasibility
of the comparatively straightforward manipulations with Feynman
diagrams. Often, a useful methodical guide may also be provided by
analogies between the quantized field in three dimensions and its
``zero-dimensional" formal analogues appearing in quantum
mechanics. In suitable units the Lagrangeans are then replaced by
Hamiltonians,
 \ben
 H = H(\lambda)= -\frac{d^2}{dx^2} + m^2 x^2
 +\lambda\,\left (
 g_3\,x^3+g_4\,x^4 + \ldots
 \right ), \ \ \ \ \ \ |\lambda| \ll 1\,
 %\label{SEb}
 \een
and the spectra and/or bound eigenstates are sought in the form of
the Rayleigh-Schr\"{o}dinger asymptotic series \cite{Fluegge},
 \ben
 E=E_n(\lambda)=(2n+1)\,\sqrt{m^2}+\lambda\,E_n^{[1]}+\lambda^2\,
 E_n^{[2]}+ \ldots\,.
 \een
Perturbation studies of various anharmonicities proved enormously
rewarding and many convergence questions emerging within the
formalism have been clarified. These studies also helped to reveal
several unexpected features of the models themselves. {\it Pars
pro toto} let us mention that Bender and Wu \cite{Wu} revealed
that for the quartic anharmonic oscillator  with $H_{\rm
int}=\lambda\, g_4\,x^4$, {\em all} the bound-state energies
$E_n(\lambda)$ with $n=0, 1, \ldots$ coincide with the values of a
{\em single} analytic multivalued function, considered simply on
its {\em separate} Riemann sheets.

The validity of similar observations has been confirmed for a few
other types of interaction terms. During their studies an amazing
empirical observation has been made in the case of the ``first
nontrivial" cubic anharmonicity $H_{\rm int}=\lambda\, g_3\,x^3$.
An analytic continuation of its energies $E_n(\lambda)$ to the
purely imaginary coupling range $\lambda \,g_3\equiv {\rm i}\,f$
with real $f$ (where the Hamiltonian itself remains manifestly
non-Hermitian, $H \neq H^\dagger$) revealed that {\em all} the
spectrum seems to remain {\em real}, $E_n(\lambda) \in I\!\!R$.
After an original, purely perturbative support of such an
unexpected possibility given in ref.~\cite{Caliceti}, decisive WKB
and numerical demonstrations of its plausibility have been
presented in refs.~\cite{Alvarez} and \cite{BB}. Finally, rigorous
proof of the reality of the whole spectrum has been delivered in
refs.~\cite{DDT} and \cite{Shin}.

In what follows we intend to point out that the list of the
exciting features of the Hamiltonians
 \ben
 H = H(m^2,f)= -\frac{d^2}{dx^2} + m^2 x^2
 +{\rm i}\,f\,x^3
 \,
 %\label{SEb}
 \een
may be complemented by an observation concerning an interesting
nonperturbative symmetry aspect of this model.

\subsubsection*{Theorem.}

Hamiltonians $H_+=H(m^2,f)+2f^{-2}(m^2/3)^3$ and
$H_-=H(-m^2,f)+2f^{-2}(-m^2/3)^3$  are isospectral.

\subsubsection*{Proof.}

The real-mass anharmonic-oscillator Schr\"{o}dinger equation
 \be
  -\frac{d^2}{dx^2} \,\psi(x)
 + m^2 x^2 \,\psi(x)
 +{\rm i}\,f\,x^3 \,\psi(x)=E_{[RM]}\, \psi(x)
 \,,
 \ \ \ \ \ \
  \,\psi(x) \in L_2(-\infty,\infty)\,
 \label{hermi}
 \ee
defines wave functions which are analytic in the whole complex
plane. By construction they asymptotically vanish not only along
the right and left half-axes but rather within the whole
neighboring wedges, i.e., along all the half-lines at angles
$\alpha \in (-3\pi/10,\pi/10)$ (forming the right wedge) and,
similarly, within the left wedge with $\alpha \in
(9\pi/10,13\pi/10)$.

For this reason all the wave functions may be analytically
continued to a shifted integration contour $x={T}+{\rm i}\,\gamma$
at any real constant $\gamma$. We may then define the new wave
functions
 \ben
 \psi({T}+{\rm i}\,\gamma)
 \equiv
 \phi({T})\in L_2(-\infty,\infty)\,
 %\label{hermio}
 \een
which satisfy the following ``imaginary mass" Schr\"{o}dinger
equation
 \be
  -\frac{d^2}{d{T}^2} \,\phi({T})
 - m^2 {T}^2 \,\phi({T})
 +{\rm i}\,f\,{T}^3 \,\phi({T})=
 E_{[IM]}
 \, \phi({T})\,.
 \label{hermiona}
 \ee
The new and old energies are related by the identity
 \be
 E_{[IM]}\ \equiv\
 \left (E_{[RM]}+\frac{4\,m^6}{27\,f^2}
 \right )
 \,.
 \label{hermionace}
 \ee
This means that Schr\"{o}dinger equations with potentials
 \be
 V_\pm(x)=\frac{2\,\left (\pm m^2\right )^3}{27\,f^2}
 \pm m^2
 {x}^2
 +{\rm i}\,f\,{x}^3\,
 \label{hermionajde}
 \ee
are isospectral. QED.

 %\noindent
One of the puzzling consequences of the latter observation is that
perturbation theory in its standard form employing the smallness
of the coupling $f$ can only be applied in the real-mass case. The
behaviour of the isospectral partner (\ref{hermiona}) of the
real-mass problem (\ref{hermi}) becomes manifestly
non-perturbative, characterized by the divergence of the energy
shift in eq.~(\ref{hermionace}).

Although this shift smoothly disappears in the vanishing-mass
limit, it makes an impression of having a ``wrong" sign at any
$m^2 > 0$. This may be read as another paradox related to the
nontriviality of the ``correct" and positive definite scalar
product which makes our Hamiltonian physical, i.e.,
(quasi-)Hermitian~\cite{Geyer}. Indeed, only with respect to this
nonstandard scalar product (discussed and even constructed by
several authors \cite{BB}) one can understand that and how the
transition to the ``wrong sign" mass term in eq.~(\ref{hermiona})
becomes {\em precisely} compensated by the shift of the spectrum
via eq.~(\ref{hermionace}).

\vspace{5mm}
\newpage
\titlepage

\vspace{5mm}

\section*{Acknowledgement}

Work supported by the Institutional Research Plan AV0Z10480505, by
the M\v{S}MT ``Doppler Institute" project Nr. LC06002 and by
GA\v{C}R, grant Nr. 202/07/1307.

\end{document}